\documentclass[11pt,a4paper]{article}
\usepackage{graphicx}
\usepackage{authblk}
\usepackage{amsmath}
\usepackage{geometry}
\usepackage{amssymb}
\usepackage{graphicx}
\usepackage{subcaption}
\graphicspath{ {./figures/} }
\title{The ePIC Silicon Vertex Tracker: Design and Status}

\author{R. Turrisi\\on behalf of the ePIC SVT Detector Subsystem Collaboration}
\affil{INFN, Sezione di Padova, Padova, Italy}

\date{}

\begin{document}
\maketitle

\begin{abstract}
The ePIC collaboration is developing a multidetector system to explore the fundamental properties of the strong
interaction at the future Electron-Ion Collider (EIC), to be built at Brookhaven National Laboratory. A key component of the
ePIC detector is the Silicon Vertex Tracker (SVT), which provides high-precision tracking and microvertex
reconstruction. The SVT consists of the Inner Barrel (IB), the Outer Barrel (OB), and the Forward/Backward Disks, all based
on Monolithic Active Pixel Sensors (MAPS) that combine high granularity, low power consumption, and minimal material
budget. This paper presents a concise overview of the SVT design and its development status.
\end{abstract}

\section{Introduction and general description}
The Electron-Ion Collider (EIC)\cite{CDR}, the future facility at the Brookhaven National Laboratory (Upton, NY, USA),
will enable precision studies of the partonic structure of nucleons and nuclei across a broad range of Bjorken-x and
four-momentum transfer squared Q$^2$. It will operate at center-of-mass energies from 28 to 140~GeV, high luminosity
(10$^{34}$cm$^{-2}$s$^{-1}$) and the unique capability to collide polarized beams (electrons, protons, and light ions) with polarization up to
$P\sim$70\% (see \cite{WP,YR}) \\ The ePIC (electron-Proton/Ion Collider Collaboration) detector has been conceived to fully exploit the EIC
physics potential (see e.g. \cite{SDT}). Its broad range of observables demands excellent performance in key measurements such as
scattered--electron kinematics, hadron reconstruction, and microvertexing, the latter being essential for the detection of heavy flavor
weak decays (c$\tau\sim$100 $\mu$m).  To meet these requirements, ePIC implements a compact and hermetic detector concept
with strong capabilities in tracking, calorimetry, and particle identification. Operations are expected to begin between
mid--2034 and early 2035.
 
The Silicon Vertex Tracker is the innermost subsystem of the tracking system. It provides precise reconstruction of
primary and secondary vertices (at the level of hundreds of microns) and contributes essential information for momentum
determination (see e.g.~\cite{laura1}). The available detector envelope constrains the SVT to $-105 < z < 135$~cm and a maximum radius of
approximately 50~cm. Combined with the 1.7~T solenoidal field, this leads to target performance requirements of 25~$\mu$m resolution on the 
single track distance of closest approach to the vertex and 0.5\% transverse momentum resolution at $p_T=1$~GeV$/c$ in the central region.

The chosen technology is MAPS, produced in a 65~nm commercial CMOS process, 
in the MOSAIX implementation developed by the ALICE Collaboration for the ITS3 upgrade~\cite{its3tdr}. 
The MOSAIX design employs wafer-scale sensors produced via stitching,
forming rows (''segments'') of 12 repeated sensor units (RSUs, groups of pixels arrays) over a length of about 266~mm. 
The sensors are thinned to 50~$\mu$m to allow bending into cylindrical shapes and to reduce multiple scattering. 
The intrinsic spatial resolution is estimated to be 7~$\mu$m with a pixel size of $21\times23~\mu\mathrm{m}^2$.  

The SVT consists of the Inner Barrel (IB), the Outer Barrel (OB), covering the central pseudorapidity range, and ten end--cap disks,
five per side, for a total active area of approximately 8.5 m$^2$.  
\\ 
IB sensors will be three, four and five segments wide for L0, L1 and L2, respectively, each sensor diced as a single piece from the wafer.  
The sensors will be placed side-by-side 
and bent into a cylindrical shape at the corresponding radius: two wafer-scale sensors on L0 and L1, and four for 
L2, with bending expertise partly inherited from ITS3. Each segment has a Left and Right Endcap (LEC, REC) for powering and
readout the former, for powering only the latter.  
The LEC region (first 4.5 mm) dissipates up to 1.6 W/$cm^2$, dominating the local cooling requirements.  For the
OB and disks, an edited MOSAIX design called the EIC Large Area Sensor (EIC-LAS) is being developed, using 5-- or 6--chip segments,
reducing design risks and improving production yield. Because the EIC-LAS calls for a different design of power
distribution, slow control, and back-biasing schemes, an external ASIC (AncASIC) implemented in the XFAB XT011 SOI process
will support these functions, reducing the power density of the LEC to 0.72~W/cm$^2$.

Overall, the SVT covers about 8~m$^2$ of silicon active area and offers full azimuthal coverage for $|\eta|\lesssim 0.9$ and
extends to $-1.64<\eta<1.88$ thanks to the endcaps. The material budget is 0.07\%~$X/X_0$ per layer in the IB, 0.25\% in
OB Layer~3 and the ten disks, and 0.55\% in OB Layer~4.
%
%
\section{Inner Barrel}
\begin{figure}[ht]
\centering
\begin{minipage}[b]{0.8\textwidth}
\begin{minipage}[b]{0.4\textwidth}
    \includegraphics[width=\textwidth]{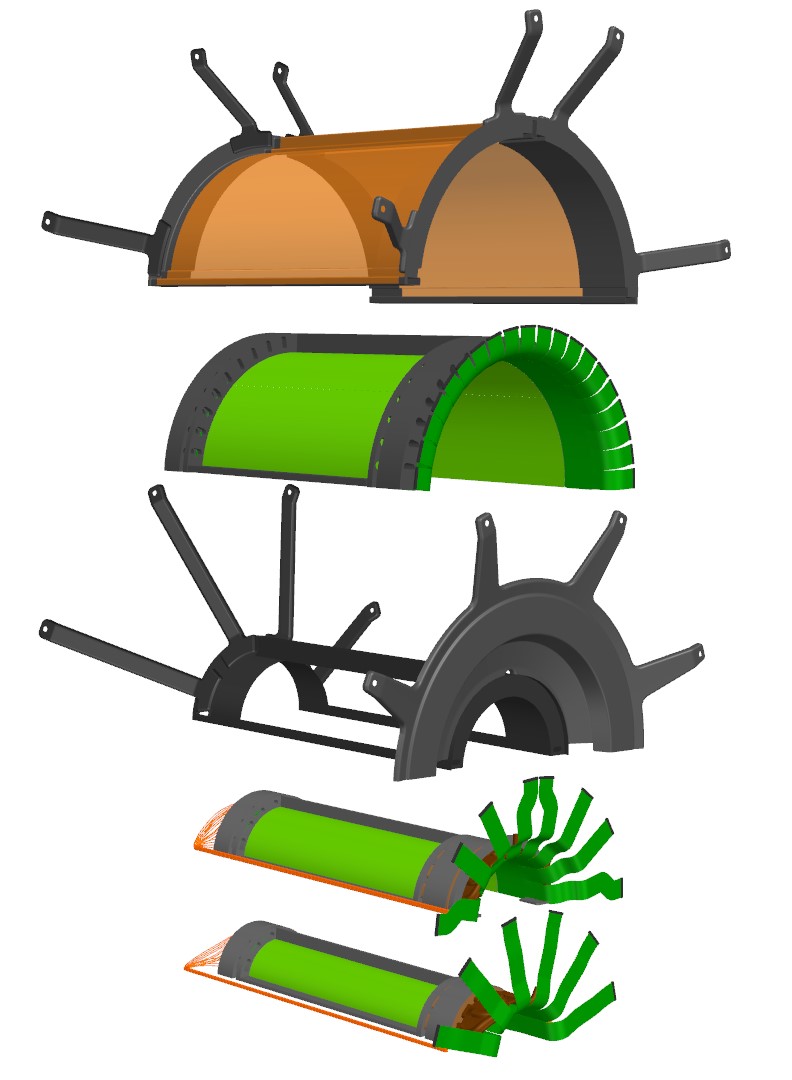}
\end{minipage}%
\hfill
\begin{minipage}[b]{0.495\textwidth}
    \includegraphics[width=\textwidth]{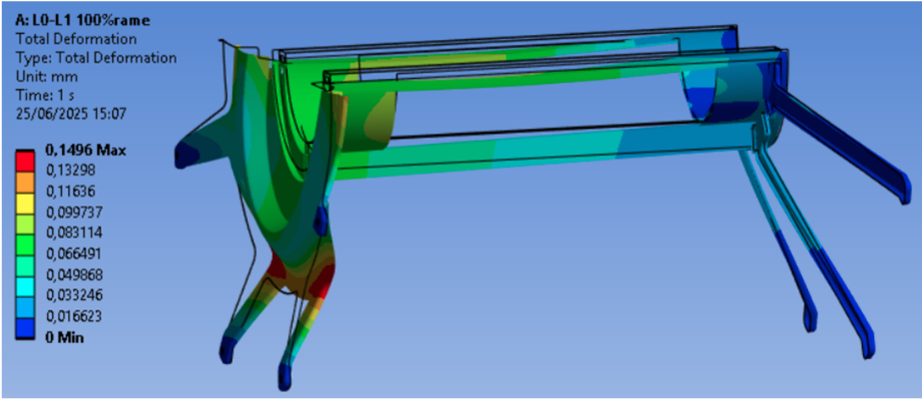}
    \vfill
    \includegraphics[width=\textwidth]{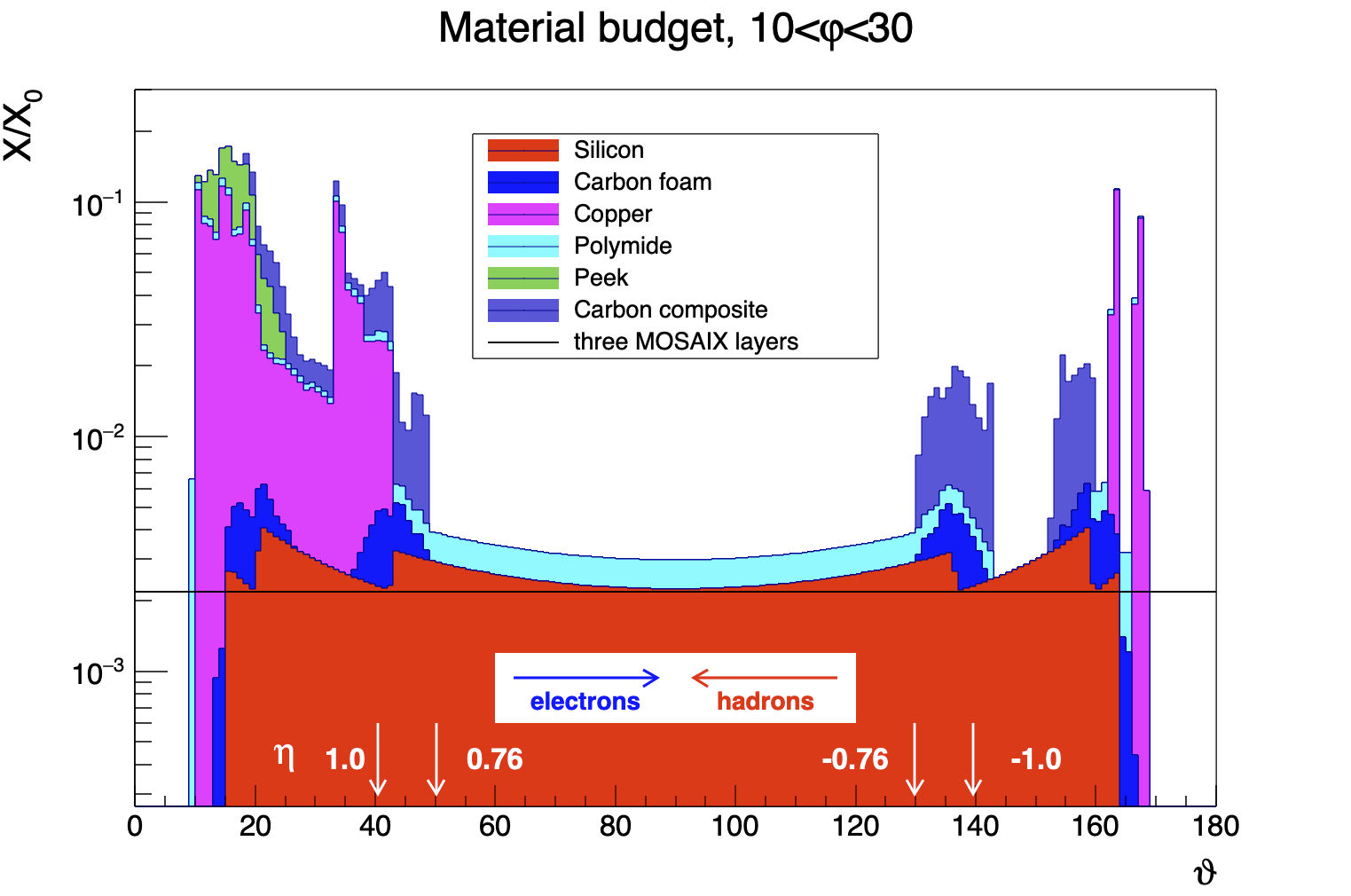}
\end{minipage}
\end{minipage}
\caption{Left panel: CAD exploded view of IB. Sensors and FPCs are in green, local mechanics in light grey, CFC support in grey, polyimide layer is in orange, and same color for the voltage supply cables on the ''electron''side. Right panel, top: mechanical load simulation.  A safety factor of 1.5 has been used on the load, FPCs are in copper. Bottom: distribution of material budget as a function of angle. Azimuth is  chosen in order to avoid the horizontal 
beams of the support.}
\label{IBCAD}
\end{figure}

The current IB design foresees two cylindrical frame structures for L0-L1 and L2, to place the sensors at radii of 38, 50 and 125 mm respectively, 
made of two symmetric half-layers. Sensors are supported by local structures made of carbon foam,  Flexible Printed Circuits (FPCs)  are 
wire-bonded to the sensor peripheries for powering and data/control transmission. The CAD exploded view of the IB design is shown in Figure~ \ref{IBCAD}.
The low power consumption of MOSAIX enables air cooling, minimising services and support material.
As shown in Figure~\ref{IBCAD}, right--bottom panel, for $|\eta|\leq 0.76$, $|\eta|\leq0.76$, only sensor material contributes significantly to the material budget; 
elsewhere, FPCs dominate. Aluminum-based FPCs are under development to further reduce mass.
This model considers an upper safe estimate value for composite support thickness of 1 mm and a simplified L2 design (still under development).  

FEA for the mechanical load (using Siemens NX\texttrademark) are carried out with a load safety factor of 1.5 and L2 deformation results are extrapolated by L0–L1. The current design presents an enhanced deformation on the hadron-side arms highlighted in orange and red, which could generate a
displacement of the support dangerous for the sensors (Figure~\ref{IBCAD}, top--right panel). Benchmark of this result by experimental tests on prototypes is 
mandatory to apply the proper modification to the design. 

\begin{figure}[ht]
\centering
\begin{minipage}[b]{0.9\textwidth}
\includegraphics[width=0.49\textwidth]{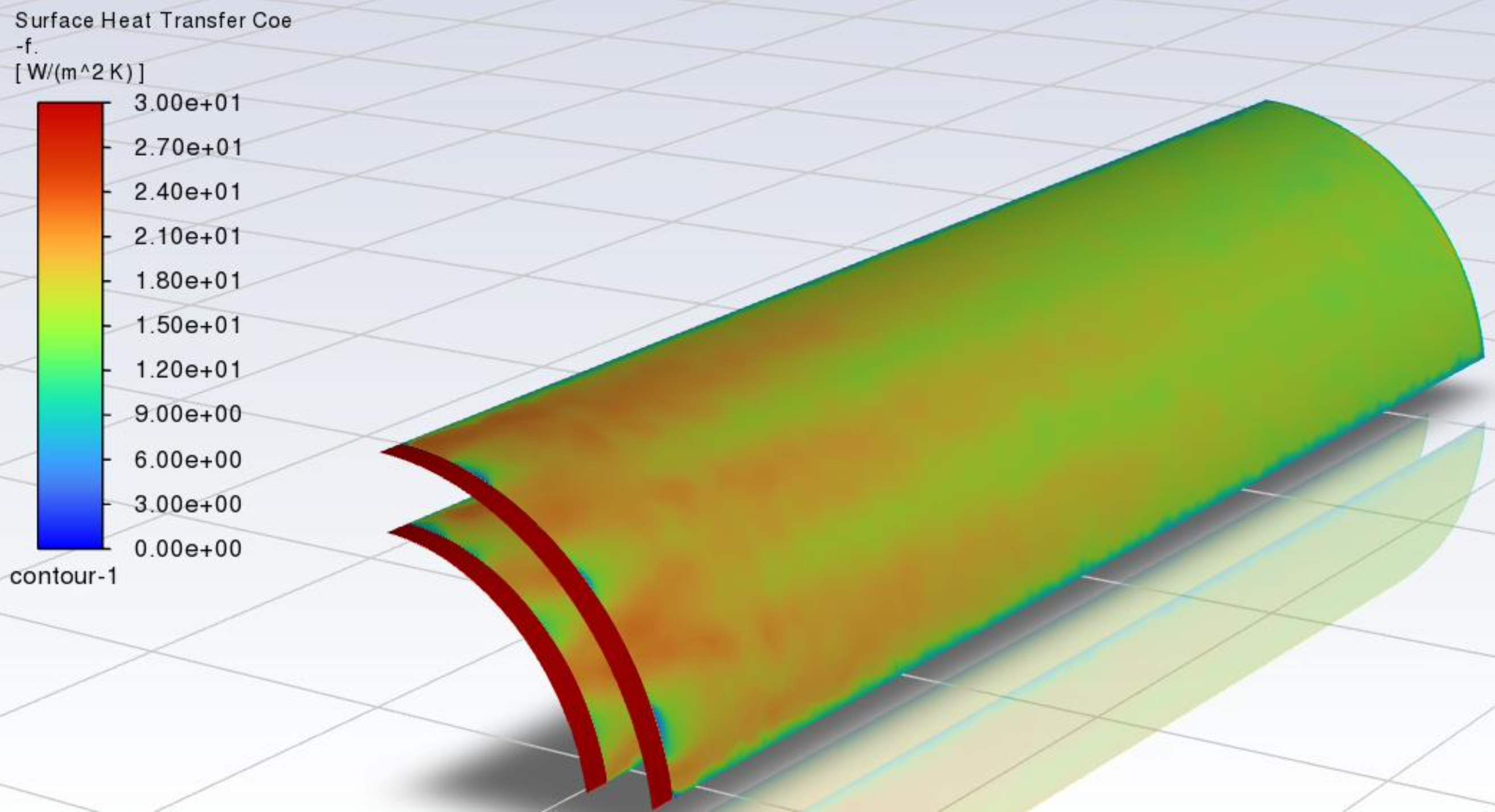}\hfill
\includegraphics[width=0.49\textwidth]{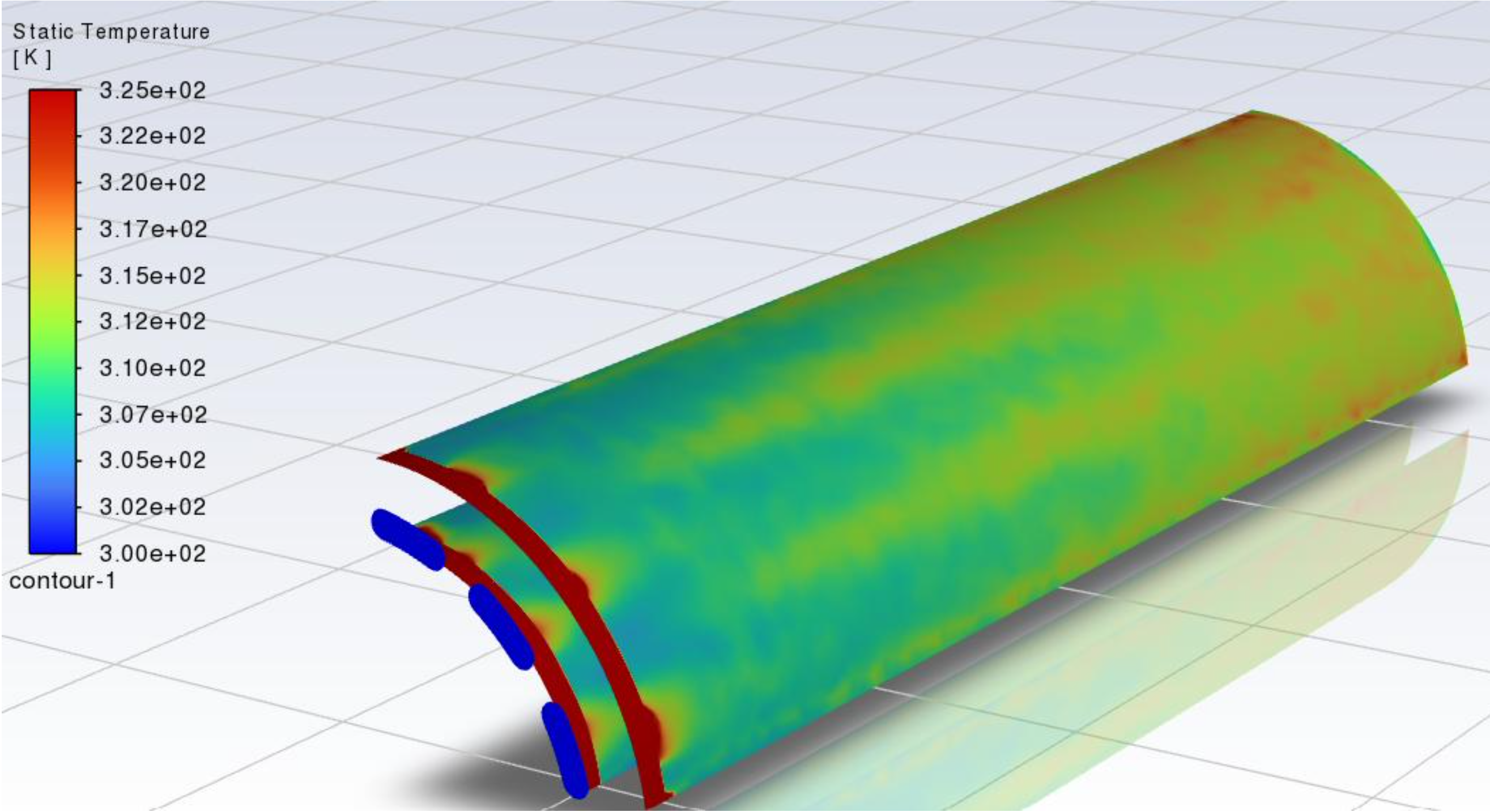}
\caption{Left: Simulation of the heat transfer from a surface at the nominal power dissipation of the MOSAIX sensor to the flowing air. Left: map of the heat transfer coefficient. Right: temperature map on the heating surface.}
\label{fluentpd}
\end{minipage}
\end{figure}

\begin{figure}[ht]
\centering
\includegraphics[width=0.35\textwidth]{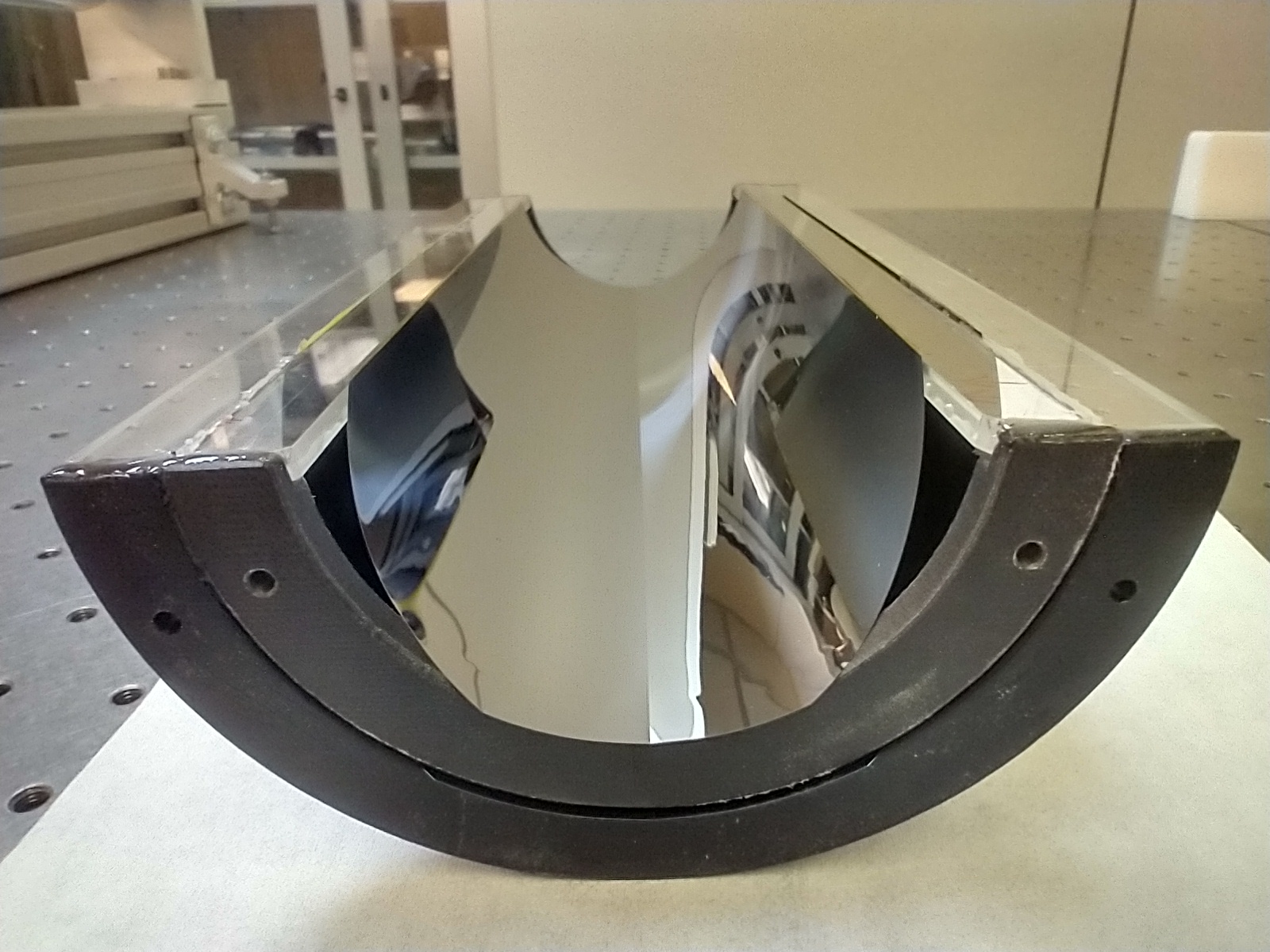}
\caption{Assembly of L0+L1  half-layers with bare silicon cutouts and 3D-printed local mechanics.}
\label{baribe}
\end{figure}

Preliminary thermal finite-element analyses by Ansys Fluent\texttrademark\  of quarter-staves suggest a heat exchange coefficient of about 10~W/m$^2$K. Temperatures in the LEC region can reach $\sim50^\circ$C in preliminary models, indicating that improved turbulence set-up and alternative 
cooling strategies for the LEC may be required, see Figure~\ref{fluentpd}, top--right panel.
Periodically repeated tests of bending of L0+L1, using  bare-silicon cutouts and 3D-printed polylactide local mechanical supports guided the tools design refinement. In Figure~\ref{baribe} the first successful assembly is shown. The same procedure will soon be extended with the integration on the IB mechanical support prototype which will be available by the end of the year.
\section{Outer Barrel}
\begin{figure}[ht]
\centering
\begin{minipage}[b]{0.4\textwidth}
    \includegraphics[width=\textwidth]{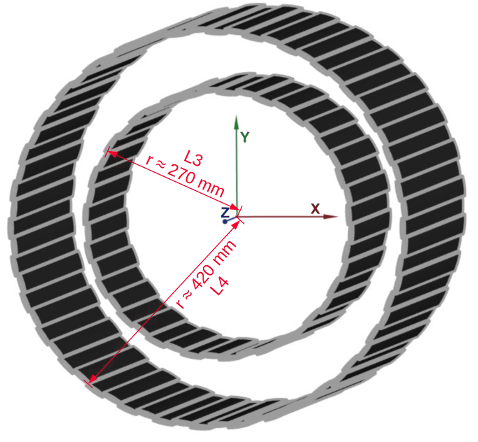}
\end{minipage}
\hfill
\begin{minipage}[b]{0.55\textwidth}
\centering
    \includegraphics[width=\textwidth]{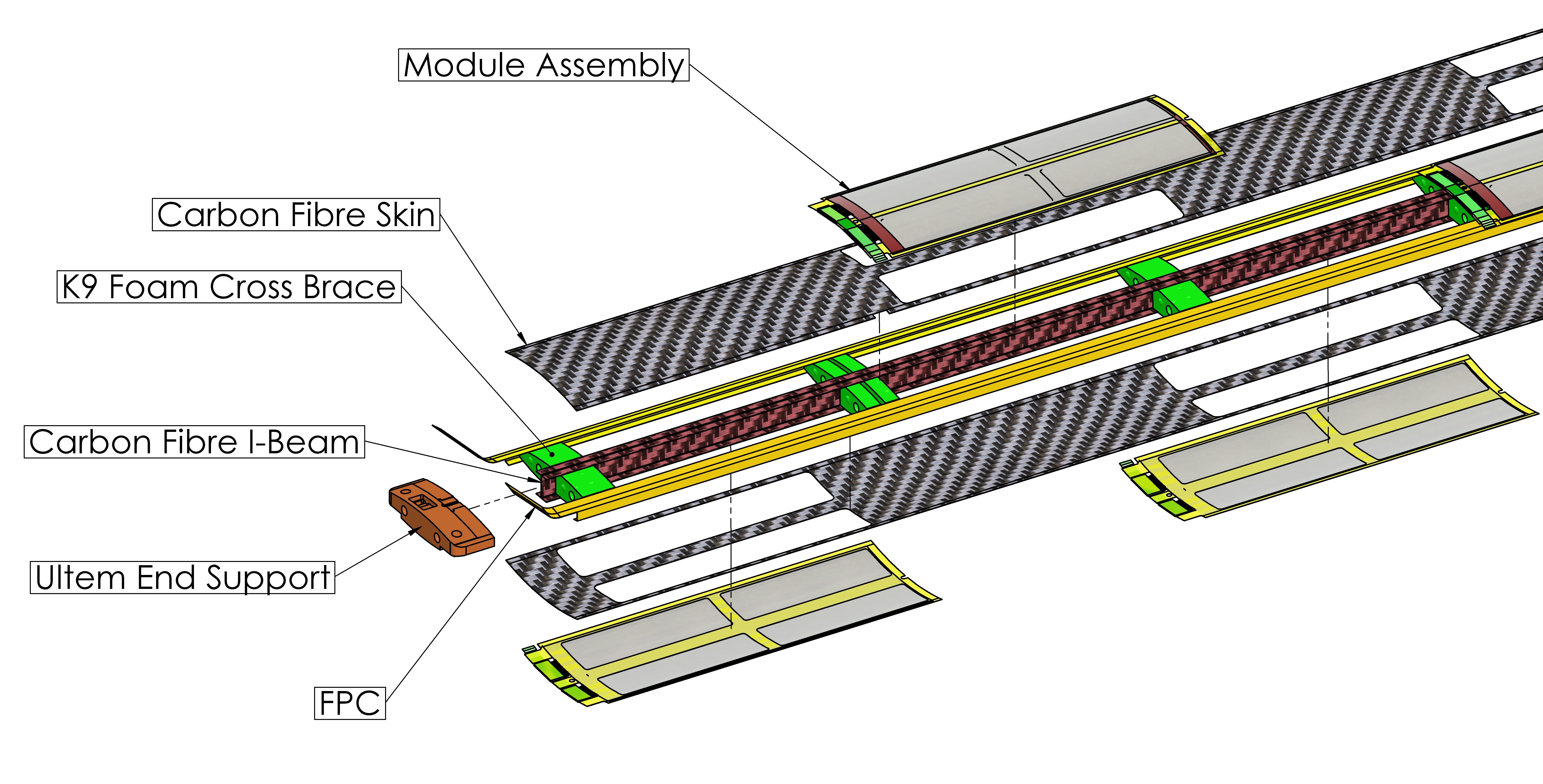}
\end{minipage}
\caption{Left: CAD representation of the Outer Barrel. Right: Exploded view of one OB module.}
\label{OB1}
\end{figure}

\begin{figure}[ht]
\begin{minipage}[b]{0.9\textwidth}
\centering
\begin{minipage}[b]{0.46\textwidth}
    \includegraphics[width=\textwidth]{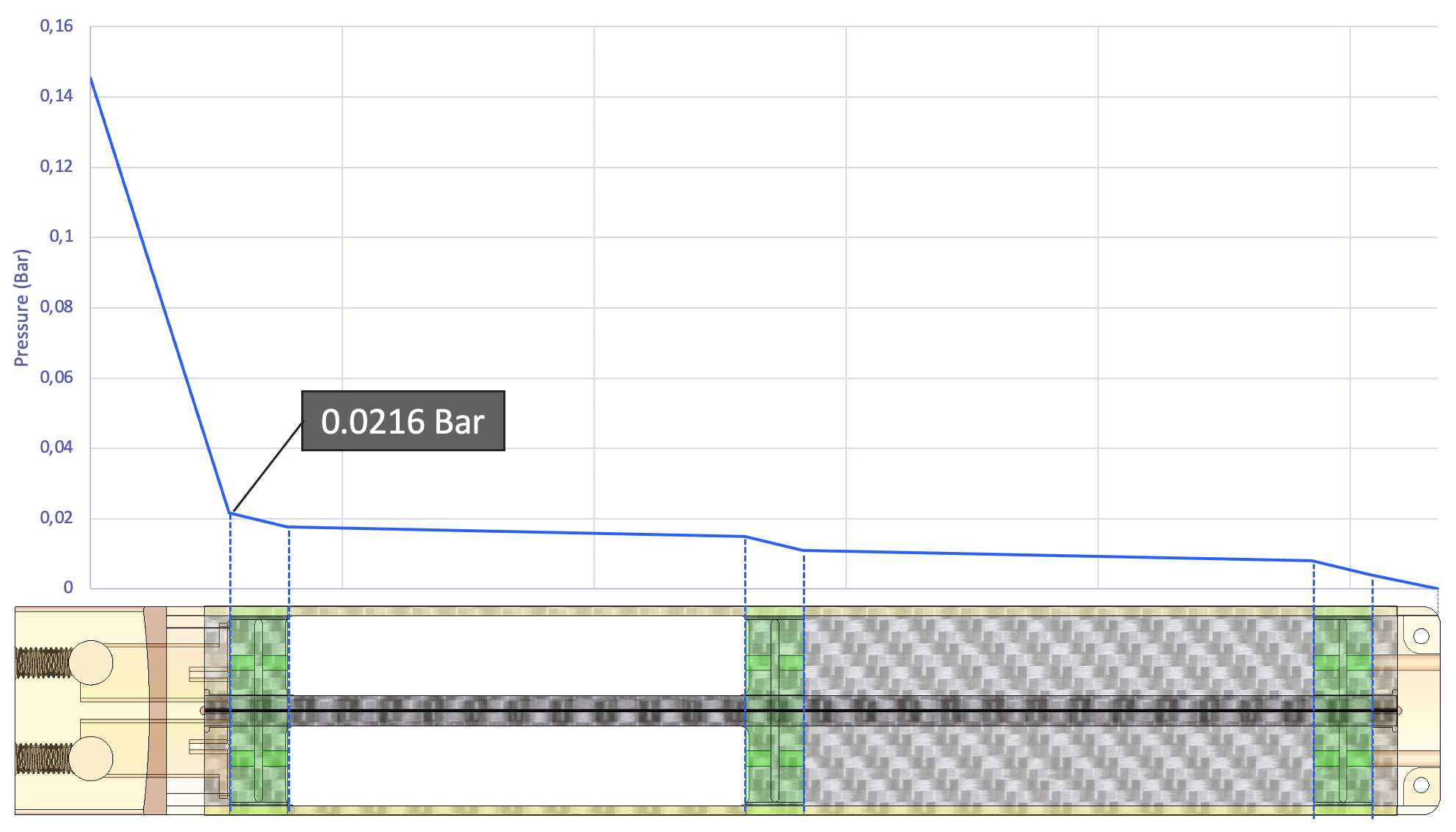} 
\end{minipage}\hfill
\begin{minipage}[b]{0.46\textwidth}
    \includegraphics[width=\textwidth]{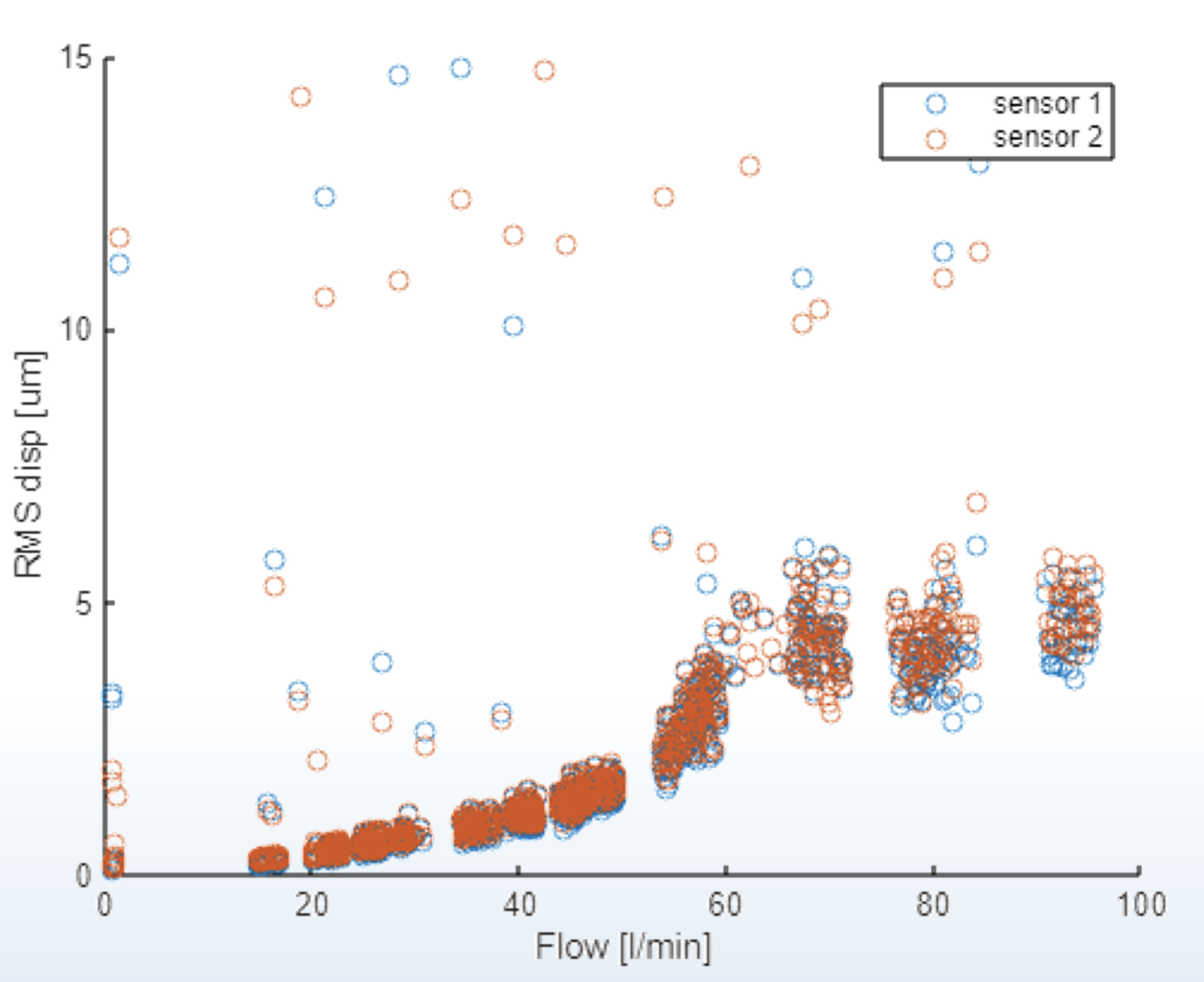}
\end{minipage}
\end{minipage}
\caption{Left: Air pressure drop measured across the main elements of a module. Right: Measurement of displacement as a function of the air flow magnitude.}
\label{OB2}
\end{figure}

The Outer Barrel (OB) will provide high precision position measurements with large lever arm delivering the required momentum
resolution and acceptance at intermediate pseudorapidity. OB is composed of two active layers, L3 and L4, assembled into a tiled-like
barrel structure with radius of 270 and 420 mm. The layers are segmented into staves arranged 
in alternating top--bottom configuration to ensure overlap, see Figure~\ref{OB1}, left panel. 
Each stave carries two EIC-LAS sensors per module, with modules on both facings, as can be seen in Figure~\ref{OB1}, right panel, 
L3 and L4 carrying four and eight modules.
 In parallel, two L4 quarter staves prototypes have been produced, comprehensive of the carbon fibre top/bottom skins, pure Kapton
FPC mock-ups and SLA 3D-printed stave-end supports. They show no noticeable twists caused by the manufacturing process, but
reinforcements are needed because of end support deformation. 
The behavior under vibration is also studied with ANSYS Modal\texttrademark  ~simulation on a cantilevered, i.e. a diving
board, quarter-stave without sensors: the FEA gives a frequency of 97 Hz for the first mode. This configuration gives similar
magnitude frequencies of a fully supported L4 stave. The structure of the module implies an impedance for the air flow.  The assessment of the air flow parameters are mandatory to ensure proper cooling power delivery along the staves. As an example, The pressure drop has been measured, isolating the single components, and is shown in Figure~\ref{OB2}, left panel. The vibrations induced from the air flowing inside the structure of the modules are 
potentially dangerous for the sensors and the stability in time of the modules. An example of RMS displacement measurement is given 
in the right panel of Figure~\ref{OB2}, where the average in the worst region, above 60 l/m, is $\approx$ 4 $\mu$m.
Conclusive tests are expected on full-length prototypes. 
\section{Disks}
\begin{figure}[ht]
\centering
\includegraphics[width=0.4\textwidth]{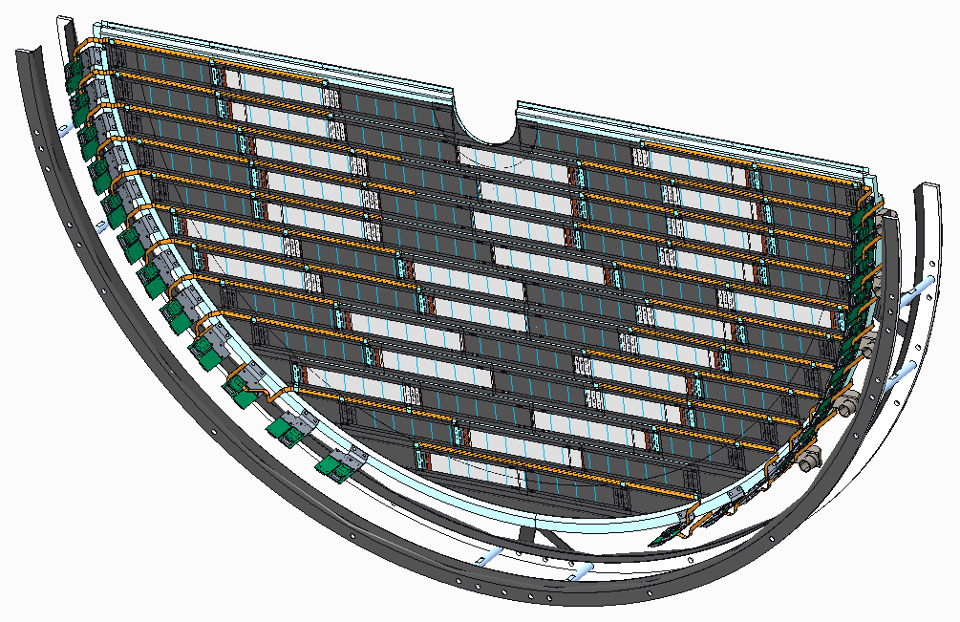}
\caption{CAD representation of a half Disk.}
\label{diskmod}
\end{figure}

\begin{figure}[ht]
\centering
\begin{minipage}[b]{0.47\textwidth}
     \includegraphics[width=\textwidth]{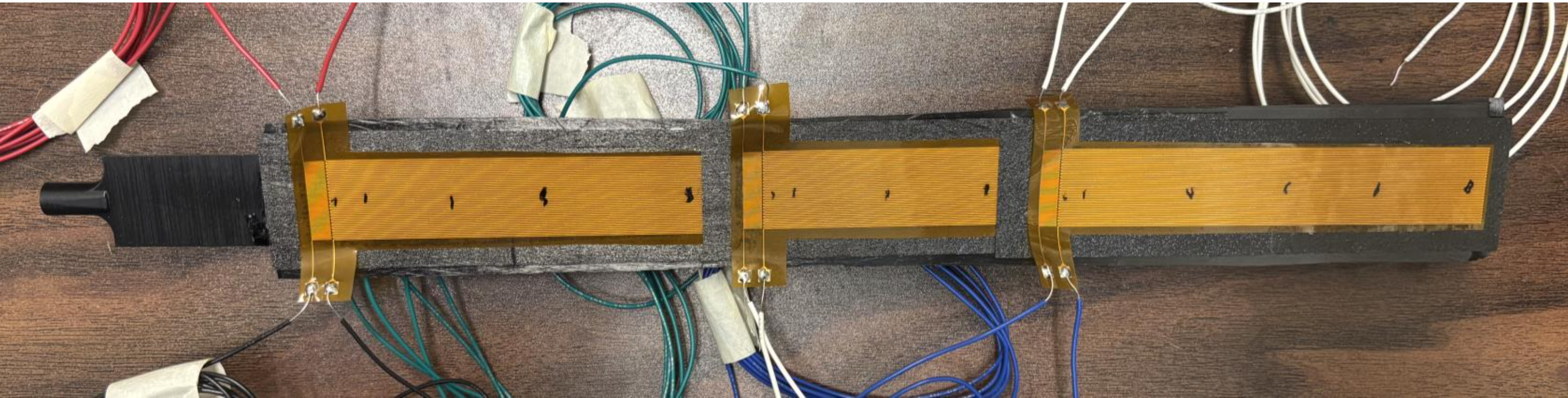}
    \vskip 10pt
    \includegraphics[width=\textwidth]{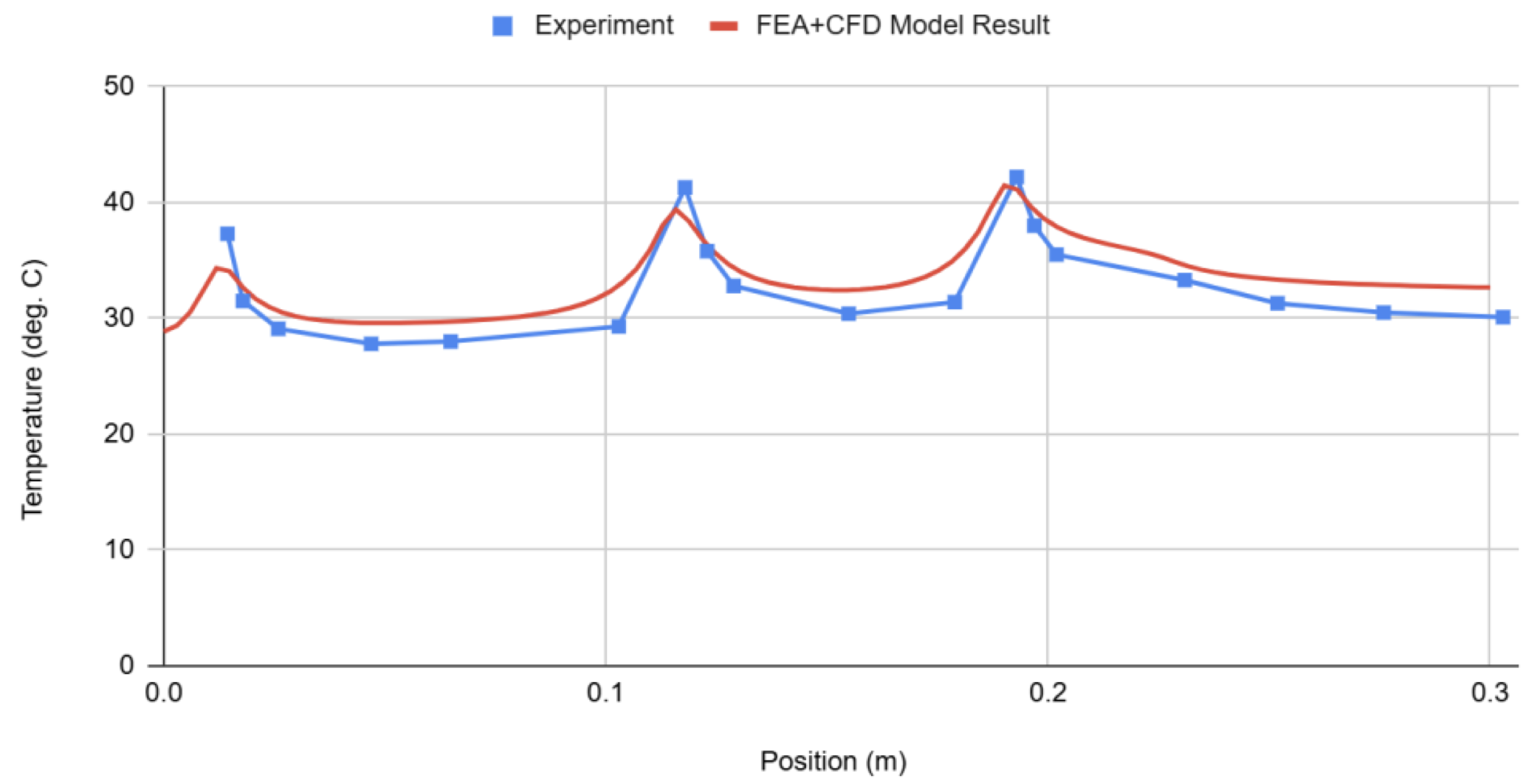}
\end{minipage}%
\hfill
\begin{minipage}[b]{0.503\textwidth}
\includegraphics[width=\textwidth]{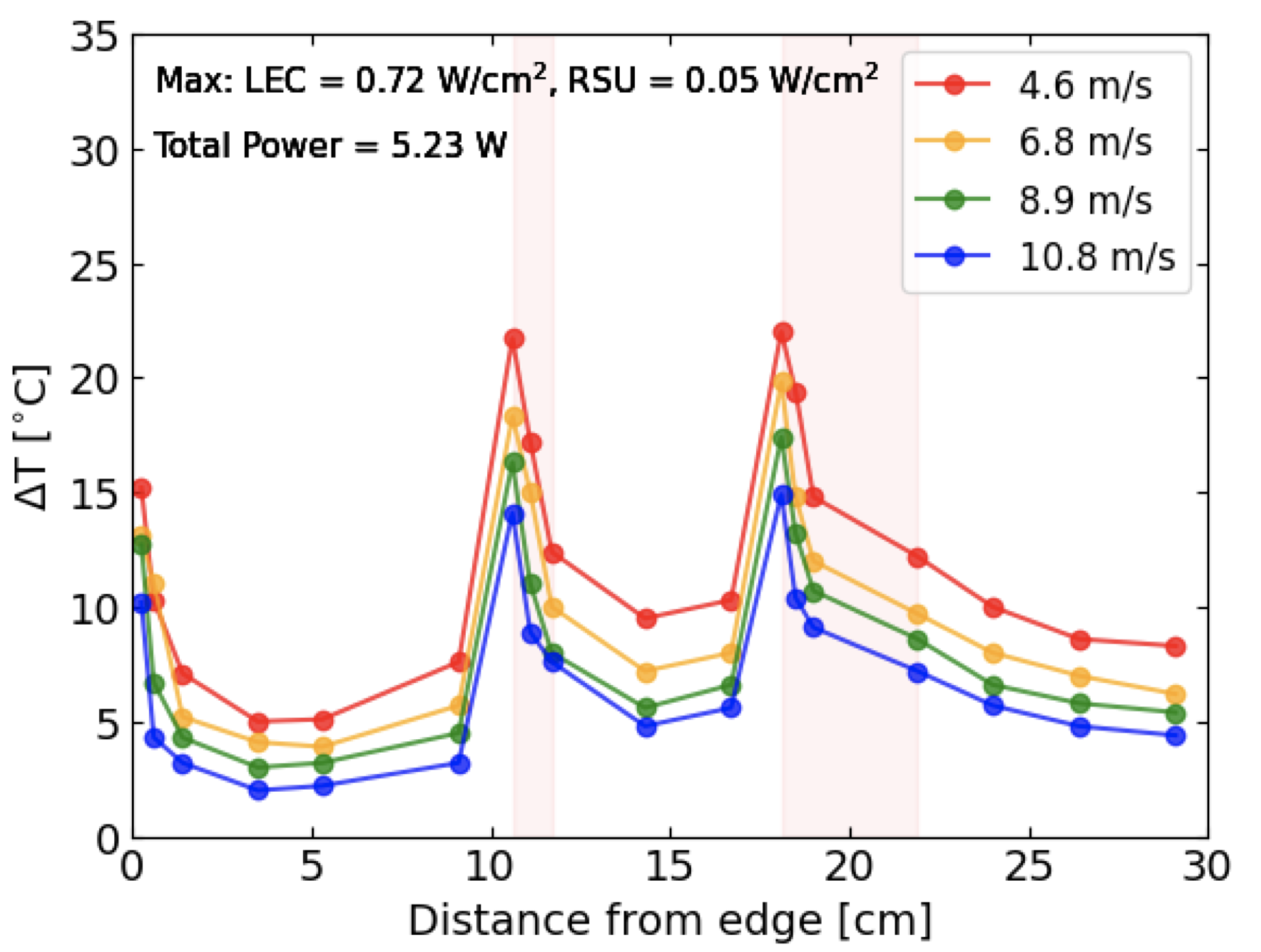}
\end{minipage}
\caption{Top-left: thermal prototype of a subset of disks module. Bottom-left: comparison of FEA+CFD model calculations with the measured temperature profile. Right: thermal profiles of the same setup obtained at four different air speeds.}
\label{disktherm}
\end{figure}

The forward and backward rapidity intervals $\eta \in$ [-1.64,-0.91] and [0.91,1.88] are instrumented with 
 five double-sided corrugated carbon-composite disks per side, assembled in halves, as shown in Figure~\ref{diskmod}. 
Outer radii range from approximately 240 to 420~mm, while inner radii increase with distance from the interaction point to accommodate the 
diverging beam pipes. 
EIC-LAS sensors are arranged with variable overlap to ensure hermeticity and optimized material distribution, 
leaving the larger values to the outer regions. Two different types of sensor modules are considered, in order to allow for a optimal assembly with inward- and outward-facing modules.
A prototype of corrugated disk support was produced using three layers of K13C2U high thermal conductivity carbon fiber composite in two layup configurations (0--90--0 and 90--0--90) where the numbers refer to the angle with respect to the corrugation.
0--90--0 offers thermal and mechanical advantage on the flat sheet, while 90--0--90 offers better stiffness for the corrugation.  
Thermal performance tests were conducted using Kapton+copper heaters matched to the sensor power densities (40~mW/cm$^2$ in the active area, 720~mW/cm$^2$ in the LEC). 
Comparison with FEA confirms good agreement using a heat exchange coefficient of 10~W/m$^2$K.
In Figure~\ref{disktherm} the setup is shown in the top left panel with the three heaters partially overlapping as in the current design.
In the same figure, right panel, the thermal profiles of the dummy sensors at stability are reported for four air speeds, 
showing that nominal operating temperatures below 40$^\circ$C are achievable with adequate margin, with an input air temperature
of 25$^\circ$C and air speeds which are considered to be safe for the detector.  In the bottom--left panel of Figure~\ref{disktherm} the comparison of one temperature profile with FEA calculation is shown, with good agreement when a heat exchange coefficient of ~10 W/m$^2$K is set.
\section{Summary and Outlook}
The ePIC Silicon Vertex Tracker has been conceived as a low-mass, high-resolution silicon detector, which will assume a critical role in the realization 
of the EIC physics program, from the inclusive DIS measurement to exclusive heavy flavors analyses, providing high precision tracking and (micro)vertices detection capabilities. 
It has reached an advanced level of design maturity. At this stage, an intense activity of prototyping and testing 
is proceeding, to validate the materials, the assembly procedures and the cooling strategy. The first mechanical and thermal mock-ups have 
been built and show encouraging results towards the validation of the design strategies, with evolving FEA modeling which supports the 
 observed behavior and will provide a reference for design refinements. 
First wafer-scale MOSAIX sensors are expected by the end of 2025, enabling the assembly of fully functional prototypes during 2026. 

\end{document}